\documentclass[12pt]{iopart}

\usepackage{iopams}  
\usepackage{setstack}
\usepackage{bm}
\usepackage[dvipdfmx]{graphicx}
\usepackage{cite}
\begin{document}

\title[Shortcuts to adiabatic cat-state generation in bosonic Josephson junctions]{Shortcuts to adiabatic cat-state generation \\ in bosonic Josephson junctions}

\author{Takuya Hatomura}

\address{Department of Physics, The University of Tokyo, 113-8654 Tokyo, Japan}
\ead{hatomura@spin.phys.s.u-tokyo.ac.jp}
\vspace{10pt}

\begin{abstract}
We propose a quantum speedup method for adiabatic generation of cat states in bosonic Josephson junctions via shortcuts to adiabaticity. 
We apply approximated counter-diabatic driving to a bosonic Josephson junction using the Holstein-Primakoff transformation. 
In order to avoid the problem of divergence in counter-diabatic driving, we take finite-size corrections into account. 
The resulting counter-diabatic driving is well-defined over whole processes. 
Schedules of the counter-diabatic driving consist of three steps; the counter-diabatic driving in the disordered phase, smoothly and slowly approaching the critical point, and the counter-diabatic driving in the ordered phase. 
Using the counter-diabatic driving, adiabatic generation of cat states is successfully accelerated. 
The enough large quantum Fisher information ensures that generated cat states are highly entangled. 
\end{abstract}

%
%
%
%
%

%
%
\section{\label{Sec.Intro}Introduction}
Schr\"odinger's cat states, i.e., superposition of macroscopically or mesoscopically distinct states, have been extensively investigated from a viewpoint of fundamental aspects of quantum mechanics~\cite{Schroedinger1935-1,Schroedinger1935-2,Schroedinger1935-3,doi:10.1143/PTP.69.80,PhysRevLett.54.857,PhysRevLett.57.13,PhysRevLett.67.2761,PhysRevA.46.4239,PhysRevLett.89.270403,RevModPhys.75.715,PhysRevLett.95.090401,PhysRevA.74.052111,PhysRevA.81.010101,1367-2630-14-9-093039,arndt2014testing,FROWIS20152,JEONG201512,PhysRevLett.115.260404,PhysRevB.93.195127,PhysRevA.94.012105,knee2016strict,tatsuta2017conversion}, and in experiments, cat states have already been observed in various systems~\cite{Monroe1131,PhysRevLett.77.4887,sackett2000experimental,friedman2000quantum,leibfried2005creation,ourjoumtsev2007generation,deleglise2008reconstruction,Jones1166,PhysRevA.82.022330,gao2008experimental,PhysRevLett.106.130506,kirchmair2012observation,Vlastakis607,Wang1087}. 
It is also of great interest to apply highly entangled cat states, e.g., the GHZ state~\cite{PhysRevLett.65.1838} and the NOON state~\cite{doi:10.1080/0950034021000011536}, to quantum metrology, where such highly entangled states provide phase sensitivity beyond the standard quantum limit to the Heisenberg limit~\cite{PhysRevA.54.R4649,Giovannetti1330,PhysRevLett.96.010401,PhysRevLett.102.100401,giovannetti2011advances,0953-4075-45-10-103001,1751-8121-47-42-424006,pezze2016non}. 
We can estimate potential usefulness of entangled states applied to quantum metrology by calculating the quantum Fisher information~\cite{PhysRevLett.102.100401}. 

As to atomic systems, however, the number of particles, of which phase sensitivity approaches the Heisenberg limit, is still limited up to $N\sim10$ due to decoherence~\cite{PhysRevLett.106.130506,pezze2016non}. 
It is one of the challenging tasks to create truly macroscopic cat states in atomic systems. 
Bosonic Josephson junctions, which consist of two coupled Bose-Einstein condensates via Josephson coupling, have been investigated to generate macroscopic cat states~\cite{PhysRevA.57.511,PhysRevA.57.1208,PhysRevA.57.2920,PhysRevA.59.4623,PhysRevA.62.013607,PhysRevA.67.013607,PhysRevA.73.023606,0295-5075-83-6-60004,0295-5075-90-1-10005,PhysRevA.81.021604,PhysRevA.85.023611,PhysRevA.95.063609,yukawa2017fast}. 
Note that if there is no energy imbalance between two modes of Bose-Einstein condensates, bosonic Josephson junctions are equivalent to the Lipkin-Meshkov-Glick model in the collective spin expression~\cite{LIPKIN1965188,MESHKOV1965199,GLICK1965211}. 
In these studies, generation schemes are divided into two types, i.e., dynamical and adiabatic generation. 
Each type of schemes has both advantages and disadvantages. 
Dynamical generation can achieve high fidelity to cat states in relatively short time. 
However, it is difficult to freeze dynamics when cat states are obtained. 
In contrast, adiabatic generation is generally robust against noises and freezes its dynamics when generation is over. 
However, it is necessary to take infinitely long time. 
The idea of adiabatic generation of cat states in Bose-Einstein condensates was first proposed by Cirac \textit{et al.}~\cite{PhysRevA.57.1208} and adiabatic dynamics of the Lipkin-Meshkov-Glick model was well-discussed in Ref.~\cite{PhysRevB.78.104426}, which suggests that fast generation of cat states by adiabatic tracking is difficult due to the existence of the critical point where the gap closes, leading to the breakdown of adiabaticity according to the adiabatic theorem~\cite{Born1928,1950435}.

One of the possible candidates overcoming this difficulty is shortcuts to adiabaticity, where we can mimic adiabatic dynamics of a desired Hamiltonian within an arbitrary time by controlling non-adiabatic transitions~\cite{doi:10.1021/jp030708a,doi:10.1021/jp040647w,doi:10.1063/1.2992152,1751-8121-42-36-365303,PhysRevLett.104.063002,Torrontegui2013117}. 
Recently, shortcuts to adiabaticity have, both exactly and approximately, been applied to many-body systems~\cite{PhysRevA.84.031606,del2012shortcuts,PhysRevLett.109.115703,PhysRevLett.111.100502,PhysRevA.90.060301}, including the Lipkin-Meshkov-Glick model~\cite{PhysRevA.86.063623,PhysRevE.87.062117,PhysRevA.88.043647,PhysRevLett.114.177206,PhysRevA.93.023815,PhysRevA.95.012309,doi:10.7566/JPSJ.86.094002}. 
In the Lipkin-Meshkov-Glick model, difficulty to find shortcuts depends on the sign of the nonlinear interaction. 
The ground state of the anti-ferromagnetic Lipkin-Meshkov-Glick model is the spin-squeezed Dicke state, which is unique and has been successfully produced using shortcuts to adiabaticity with high fidelity and within short time~\cite{PhysRevA.86.063623,PhysRevA.88.043647,PhysRevA.93.023815}. 
In contrast, the ground state of the ferromagnetic Lipkin-Meshkov-Glick model is the cat state. 
Shortcuts to adiabaticity in the ferromagnetic Lipkin-Meshkov-Glick model was first studied by Takahashi using the Holstein-Primakoff transformation in the thermodynamic limit~\cite{PhysRevE.87.062117}. 
Counter-diabatic driving was derived for both the disordered and the ordered phases. 
However, this counter-diabatic driving is ill-defined, i.e., diverges, at the critical point unless the fixed-point condition is satisfied. 
As discussed in literatures, especially in Ref.~\cite{PhysRevLett.114.177206}, this divergence is associated with the closing of the gap and the divergence of the correlation length. 
Campbell \textit{et al.} studied counter-diabatic driving around the critical point applying various approaches, especially in combination with optimal control~\cite{PhysRevLett.114.177206}. 
By applying a small longitudinal field, which enables us to slightly avoid the critical point, mean-field prescription was applied both in the invariant-based inverse engineering approach~\cite{PhysRevA.95.012309} and the counter-diabatic driving approach~\cite{doi:10.7566/JPSJ.86.094002}. 

In this paper, we propose approximated counter-diabatic driving for bosonic Josephson junctions without energy imbalance, which is available across the critical point, using finite-size corrections in the Holstein-Primakoff transformation. 
Advantages of our method are that the counter-diabatic driving is well-defined over whole processes and that schedules of the counter-diabatic driving can be analytically obtained. 
Using our counter-diabatic driving, we can accelerate adiabatic generation of the cat state. 
Schedules of the counter-diabatic driving consist of three steps. 
The first one is the counter-diabatic driving in the disordered phase, where we aim to let the system be in the ground state. 
The second one is smoothly and slowly approaching the critical point, where we give up to be adiabatic but aim to suppress unfavorable transitions up to the first excited state which degenerates into the ground state in the thermodynamic limit. 
The last one is the counter-diabatic driving in the ordered phase. 
Improved adiabaticity will be demonstrated using the distribution of the eigenstate populations, the fidelity to the ground state subspace, the residual energy, the incomplete magnetization, and the quantum Fisher information. 
These results support usefulness of our method. 
Note that our method should be feasible in experiments using the recent theoretical proposal~\cite{PhysRevA.91.053612} and state of the art experimental techniques. 

This paper is constructed as follows. In Sec.~\ref{Sec.BJJ}, we introduce bosonic Josephson junctions without energy imbalance. 
The basic properties of bosonic Josephson junctions, which are necessary for our discussions, are reviewed. 
We summarize counter-diabatic driving for non-degenerate and degenerate eigenstates in Sec.~\ref{Sec.CDdrivingND} and Sec.~\ref{Sec.CDdrivingD}, respectively. 
The counter-diabatic driving in the Lipkin-Meshkov-Glick model using the Holstein-Primakoff transformation in the thermodynamic limit is reviewed in Sec.~\ref{Sec.CDdrivingLMG}. 
In Sec.~\ref{Sec.finite}, we develop the counter-diabatic driving in the Lipkin-Meshkov-Glick model using the Holstein-Primakoff transformation considering finite-size corrections. 
In Sec.~\ref{Sec.cat}, we discuss generation of the cat state via shortcuts. 
We summarize in Sec.~\ref{Sec.conc}.

%
%
\section{\label{Sec.BJJ}Bosonic Josephson junctions}
The Hamiltonian of a bosonic Josephson junction without energy imbalance between two modes of Bose-Einstein condensates, which is equivalent to the Lipkin-Meshkov-Glick model in the collective spin expression, is given by
\begin{equation}
\mathcal{H}_0(t)=\hbar\chi S_z^2+\hbar\Omega(t)S_x, 
\label{Eq.BJJ}
\end{equation}
where $\chi$ is the strength of the nonlinear interaction, $\Omega(t)$ is the strength of the Josephson coupling, and $S_\alpha$, $\alpha=x, y, z$, is the usual angular momentum operator~\cite{PhysRevA.55.4318,PhysRevA.57.1208,PhysRevA.57.2920,RevModPhys.73.307,0953-4075-40-10-R01,doi:10.1080/09500340.2011.632100}. 
The size of the angular momentum $\bm{S}^2=S(S+1)$ is given by $S=N/2$, where $N$ is the fixed total number of atoms. 
For simplicity, hereafter, we consider the time-independent interaction $\chi$ and the time-dependent coupling $\Omega(t)$ and we put $\hbar=1$. 
Because we are interested in cat states, we assume the negative interaction $\chi<0$, with which the ground state of the bosonic Josephson junction (\ref{Eq.BJJ}) becomes the cat state in the ordered phase~\cite{GILMORE1978189,PhysRevLett.49.478,PhysRevB.28.3955,PhysRevLett.93.237204,PhysRevB.71.224420}. 
By using the notations of the Lipkin-Meshkov-Glick model, $\chi=-2J/N$ and $\Omega(t)=-2\Gamma(t)$, in the thermodynamic limit, $N\to\infty$, the critical point at $\Gamma(t)=J$ divides into two phases, the disordered phase for $\Gamma(t)>J$ and the ordered phase for $\Gamma(t)<J$. 

\begin{figure}
\centering
\includegraphics[width=10cm]{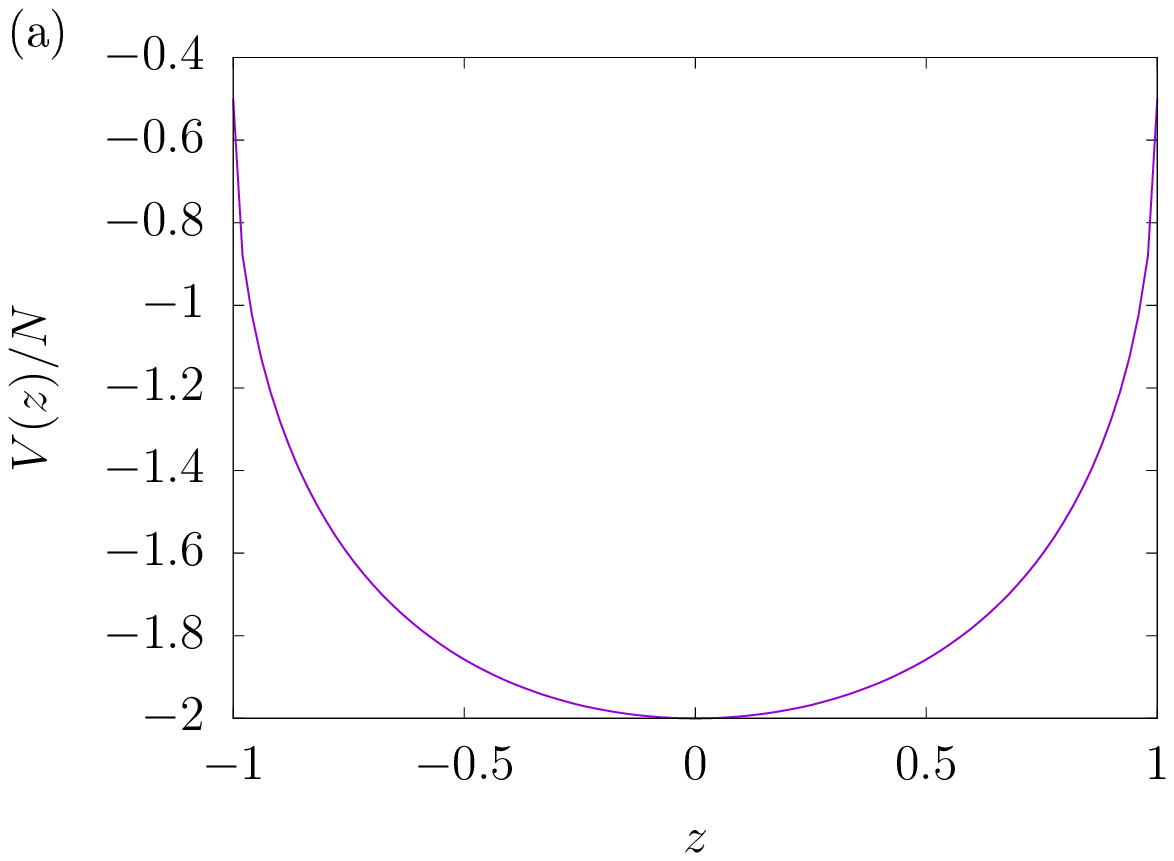}
\includegraphics[width=10cm]{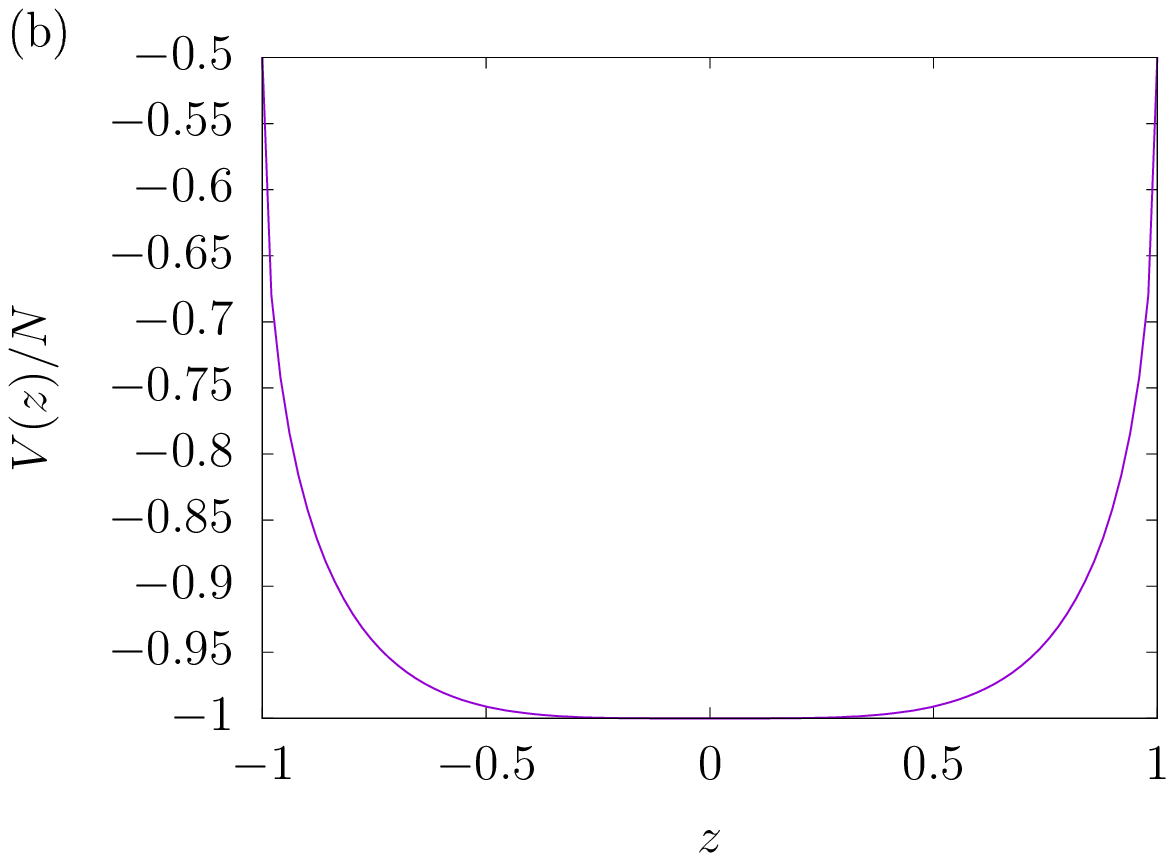}
\includegraphics[width=10cm]{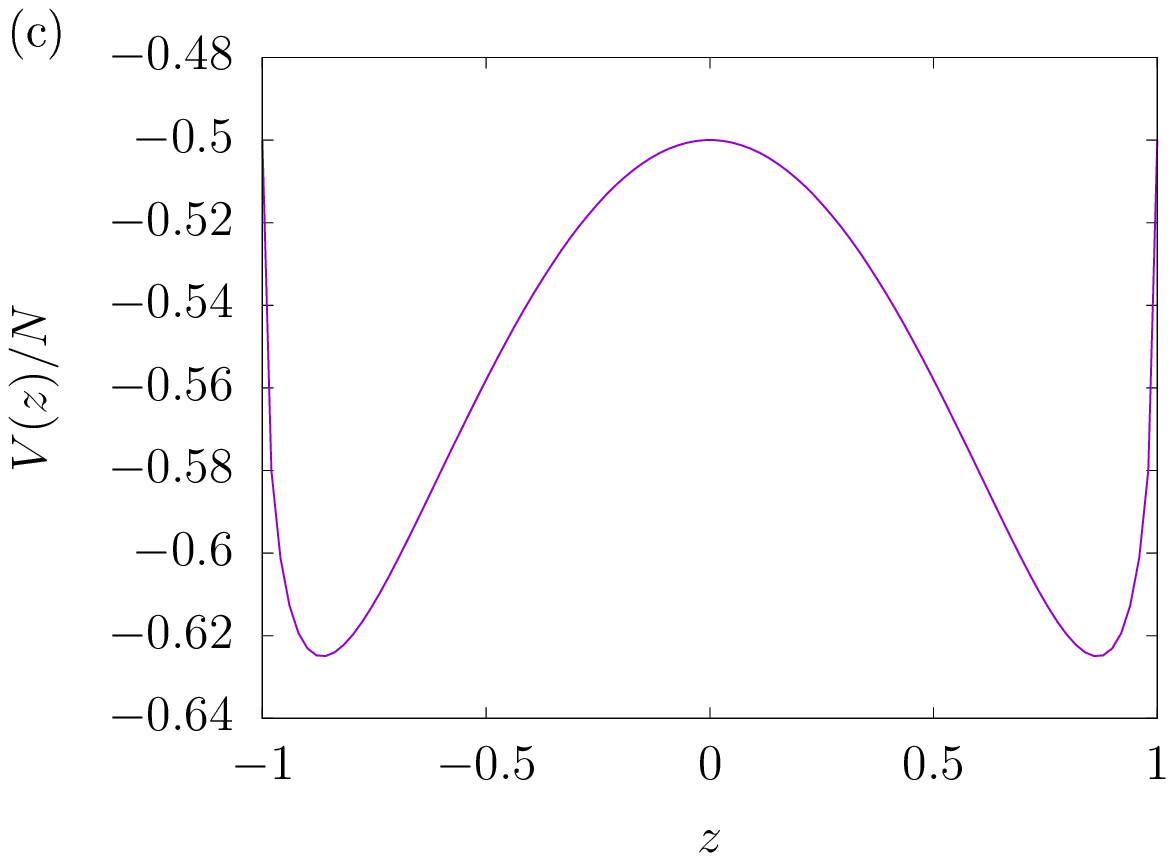}
\caption{\label{Fig.potential} Potential energy of the bosonic Josephson junction (\ref{Eq.BJJ}) for (a) $\Gamma(t)=2J$, (b) $\Gamma(t)=J$, and (c) $\Gamma(t)=0.5J$. Here, $J=1$. }
\end{figure}

It is useful to apply the semi-classical approximation in order to investigate behavior of the Hamiltonian (\ref{Eq.BJJ})~\cite{PhysRevLett.79.4950,PhysRevA.59.620,PhysRevA.78.023611,0295-5075-83-6-64007,PhysRevA.81.063625,PhysRevA.86.023615}. 
Introducing the continuous parameter $z$, limited in the range $-1\le z\le1$, and the azimuth $\theta$, we replace the angular momentum operators with
\begin{eqnarray}
&S_x=\frac{N}{2}\sqrt{1-z^2}\cos\theta, \\
&S_y=\frac{N}{2}\sqrt{1-z^2}\sin\theta, \\
&S_z=\frac{N}{2}z, 
\end{eqnarray}
and thus the Hamiltonian (\ref{Eq.BJJ}) becomes
\begin{eqnarray}
\mathcal{H}_0(t)&=&-\frac{JN}{2}z^2-\Gamma(t)N\sqrt{1-z^2}\cos\theta \\
&\simeq&\frac{\Gamma(t)N}{2}\theta\sqrt{1-z^2}\theta-\frac{JN}{2}z^2-\Gamma(t)N\sqrt{1-z^2},
\end{eqnarray}
where we assume $\theta\ll1$. 
Here, the first term corresponds to the kinetic energy and the second and the third terms represent the potential
\begin{eqnarray}
V(z)&=&-\frac{JN}{2}z^2-\Gamma(t)N\sqrt{1-z^2} \label{Eq.potential} \\
&\simeq&-\Gamma(t)N+\frac{1}{2}(\Gamma(t)-J)Nz^2+\frac{1}{8}\Gamma(t)Nz^4.
\end{eqnarray}
We plot the potential (\ref{Eq.potential}) for three regions in Fig.~\ref{Fig.potential}. 
In the disordered phase, $\Gamma(t)>J$, the ground state is distributed around $z=0$ and thus this potential can be regarded as the harmonic potential. 
However, close to the critical point, $\Gamma(t)\to J$, the quadratic term becomes small, leading to the quartic-like potential. 
In the ordered phase, $\Gamma(t)<J$, two minima appear in the quartic-like potential, and thus the harmonic approximation at each minimum is valid. 
Note that two minima are given by $z=\pm\sqrt{1-(\Gamma(t)/J)^2}$ and the altitude of the spin vector $\phi$ is given by $\cos\phi=\Gamma(t)/J$. 

%
%
\section{\label{Sec.CDdriving}Counter-diabatic driving}
%
%
\subsection{\label{Sec.CDdrivingND}Non-degenerate eigenstates}
Counter-diabatic driving in non-degenerate systems was developed by Demirplak and Rice~\cite{doi:10.1021/jp030708a,doi:10.1021/jp040647w,doi:10.1063/1.2992152} and by Berry~\cite{1751-8121-42-36-365303}, independently. 
We consider a time-dependent Hamiltonian with the non-degenerate eigenstates
\begin{equation}
\mathcal{H}_0(t)=\sum_nE_n(t)|n(t)\rangle\langle n(t)|,
\end{equation}
where $E_n(t)$ is the eigen-energy and $|n(t)\rangle$ is the eigenstate of the Hamiltonian $\mathcal{H}_0(t)$. 
If the Hamiltonian varies slowly in time, the adiabatic approximation holds and thus each energy eigenstate evolves as
\begin{equation}
|\psi_n(t)\rangle=e^{i\alpha_n(t)}|n(t)\rangle,
\end{equation}
where $\alpha_n(t)$ is usually expressed by the dynamical and the Berry phases~\cite{Berry45,PhysRevLett.51.2167}
\begin{equation}
\alpha_n(t)=-\int_0^tdt^\prime E_n(t^\prime)+i\int_0^tdt^\prime\langle n(t^\prime)|\partial_{t^\prime}n(t^\prime)\rangle.
\end{equation}
The Hamiltonian which mimics such adiabatic dynamics is given by
\begin{equation}
\mathcal{H}(t)=\mathcal{H}_0(t)+\mathcal{H}_\mathrm{cd}(t),
\label{Eq.TotHam}
\end{equation}
where $\mathcal{H}_\mathrm{cd}(t)$ is the counter-diabatic Hamiltonian
\begin{eqnarray}
\mathcal{H}_\mathrm{cd}(t)&=&i\sum_n(1-|n(t)\rangle\langle n(t)|)|\partial_tn(t)\rangle\langle n(t)| \label{Eq.CDHamND1} \\
&=&i\sum_{n\neq m}\frac{\langle m(t)|(\partial_t\mathcal{H}_0(t))|n(t)\rangle}{E_n(t)-E_m(t)}|m(t)\rangle\langle n(t)|. \label{Eq.CDHamND2}
\end{eqnarray}
%
%
\subsection{\label{Sec.CDdrivingD}Degenerate eigenstates}
In degenerate systems, counter-diabatic driving was developed by Takahashi~\cite{PhysRevE.87.062117}. 
We consider a time-dependent Hamiltonian with the degenerate eigenstates
\begin{equation}
\mathcal{H}_0(t)=\sum_{n,\mu}E_n(t)|n,\mu;t\rangle\langle n,\mu;t|,
\end{equation}
where $|n,\mu;t\rangle$ is the eigenstate of the Hamiltonian $\mathcal{H}_0(t)$ and $\mu$ is an additional index due to degeneracies. 
We assume that there is no level-crossing during time-evolution. 
In this case, adiabatic dynamics of each energy eigenstate is given by
\begin{equation}
|\psi_n(t)\rangle=\exp\left(-i\int_0^tdt^\prime E_n(t^\prime)\right)\sum_\mu c_\mu^{(n)}(t)|n,\mu;t\rangle,
\end{equation}
where $c_\mu^{(n)}(t)$ is given by
\begin{equation}
c_\mu^{(n)}(t)=\sum_{\mu^\prime}U_{\mu\mu^\prime}^{(n)}(t)c_{\mu^\prime}^{(n)}(0),
\end{equation}
and
\begin{equation}
U^{(n)}(t)=\mathcal{T}\exp\left(-i\int_0^tdt^\prime A^{(n)}(t^\prime)\right),
\end{equation}
as discussed by Wilczek and Zee~\cite{PhysRevLett.52.2111}. 
Here, $A^{(n)}(t)$ is the gauge potential
\begin{equation}
iA_{\mu\mu^\prime}^{(n)}(t)=\langle n,\mu;t|\partial_tn,\mu^\prime;t\rangle.
\end{equation}
The Hamiltonian which mimics such dynamics is given by Eq.~(\ref{Eq.TotHam}) with the counter-diabatic Hamiltonian
\begin{eqnarray}
\mathcal{H}_\mathrm{cd}(t)&=&i\sum_{n,\mu}\left(1-\sum_\nu|n,\nu;t\rangle\langle n,\nu;t|\right)|\partial_tn,\mu;t\rangle\langle n,\mu;t| \label{Eq.CDHamD1} \\
&=&i\sum_{n\neq m,\mu,\nu}\frac{\langle m,\nu;t|(\partial_t\mathcal{H}_0(t))|n,\mu;t\rangle}{E_n(t)-E_m(t)}|m,\nu;t\rangle\langle n,\mu;t|. \label{Eq.CDHamD2}
\end{eqnarray}

%
%
\subsection{\label{Sec.CDdrivingLMG}Lipkin-Meshkov-Glick model}
In this section, we review the counter-diabatic driving in the Lipkin-Meshkov-Glick model using the Holstein-Primakoff transformation~\cite{PhysRevE.87.062117}. 
We remark that in the disordered phase, $\Gamma(t)>J$, $(2S+1)$ eigenstates are separated from each other, and thus the counter-diabatic Hamiltonian is constructed by Eq.~(\ref{Eq.CDHamND1}) or Eq.~(\ref{Eq.CDHamND2}). 
In contrast, in the ordered phase, $\Gamma(t)<J$, eigenstates have 2-fold degeneracies except for the highest energy eigenstate, and thus the counter-diabatic Hamiltonian is constructed by Eq.~(\ref{Eq.CDHamD1}) or Eq.~(\ref{Eq.CDHamD2}). 
Originally, doubly degenerate ground states are the counterparts of the singlet and the triplet states. 
However, using the Holstein-Primakoff transformation and the harmonic approximation means neglecting inter-well transitions, and thus the degenerate ground states become the ground states of each well. 
Therefore, in this approximation, the counter-diabatic Hamiltonian should be
\begin{eqnarray}
\mathcal{H}_\mathrm{cd}(t)&=\sum_\mu\mathcal{H}_\mathrm{cd}^{(\mu)}(t) \\
&=i\sum_{n\neq m,\mu}\frac{\langle m,\mu;t|(\partial_t\mathcal{H}_0(t))|n,\mu;t\rangle}{E_n(t)-E_m(t)}|m,\mu;t\rangle\langle n,\mu;t|,
\label{Eq.CDhamBS}
\end{eqnarray}
that is, the summation of two counter-diabatic Hamiltonian for each well, which are formally independent of each other. 
As mentioned later, however, the counter-diabatic terms for a chosen well, say the well with positive (negative) magnetization, affect the state of the other well with negative (positive) magnetization. 
Therefore, the correctness of our approximation must be confirmed by numerical simulations. 

In the disordered phase, the spin vector is directed to the $x$-axis due to the strong external field $\Gamma(t)$. 
In contrast, in the ordered phase, the spin vector rotates, pointing at a minimum of the potential, by angles $\pm\phi$, where $\cos\phi=\Gamma(t)/J$. 
By applying the Holstein-Primakoff transformation expanded up to the second order of the boson operators, which becomes exact in the thermodynamic limit, and applying the Bogoliubov transformation, the Hamiltonian (\ref{Eq.BJJ}) is mapped to the harmonic oscillator
\begin{equation}
\mathcal{H}_0(t)=\omega(t)\left(b^\dag b+\frac{1}{2}\right),
\label{Eq.harmHam}
\end{equation}
where $b$ and $b^\dag$ are the annihilation and the creation operators obtained by the Bogoliubov transformation. 
Here, we neglect constants, which differ depending on the phases. 
The frequencies are given by
\begin{equation}
\omega(t)=2\sqrt{\Gamma(t)(\Gamma(t)-J)},
\label{Eq.HPenergyPara}
\end{equation}
in the disordered phase, and
\begin{equation}
\omega(t)=2\sqrt{J^2-\Gamma^2(t)},
\label{Eq.HPenergyFerro}
\end{equation}
for each minimum of the potential in the ordered phase. 
Note that the sign of the rotation $\phi$ does not affect the transformed Hamiltonian (\ref{Eq.harmHam}) in this approximation. 

The counter-diabatic Hamiltonian of the harmonic oscillator is already known~\cite{0953-4075-43-8-085509} and given by
\begin{equation}
\mathcal{H}_\mathrm{cd}(t)=\frac{i(\partial_t\omega(t))}{4\omega(t)}(b^2-b^{\dag2}). 
\label{Eq.CDhamHarm}
\end{equation}
In the representation of the total spin operators, the counter-diabatic Hamiltonian (\ref{Eq.CDhamHarm}) is rewritten as
\begin{equation}
\mathcal{H}_\mathrm{cd}(t)=\frac{f(t)}{N}(S_yS_z+S_zS_y). 
\label{Eq.CDhamLMG}
\end{equation}
Here, using Eqs.~(\ref{Eq.HPenergyPara}) and (\ref{Eq.HPenergyFerro}), the schedules of the counter-diabatic driving are given by
\begin{equation}
f(t)=-\frac{(2\Gamma(t)-J)(\partial_t\Gamma(t))}{4\Gamma(t)(\Gamma(t)-J)},
\label{Eq.scheDO1}
\end{equation}
in the disordered phase, and
\begin{equation}
f(t)=\frac{\Gamma(t)(\partial_t\Gamma(t))}{(J^2-\Gamma^2(t))}
\label{Eq.scheO1}
\end{equation}
in the ordered phase. 
Note that the schedule of the counter-diabatic driving in the ordered phase differs by the factor two from the previous works (see e.g. Eq.~(S-17) in Supplemental Material of Ref.~\cite{PhysRevLett.114.177206}) because we use Eq.~(\ref{Eq.CDhamBS}) instead of Eq.~(\ref{Eq.CDHamND2}), which is supported by numerical simulations. 

It is obvious that the counter-diabatic Hamiltonian diverges at the critical point $\Gamma(t)=J$. 
It is confirmed as follows. 
The expansion of the transverse field $\Gamma(t)$ around the critical point $\Gamma(t)=J$ is given by
\begin{equation}
\Gamma(t)\approx J+(\partial_t\Gamma(t))|_{\Gamma(t)=J}\delta\Gamma(t)+\mathcal{O}(\delta\Gamma^2), 
\end{equation}
and thus it leads to the divergence at the critical point
\begin{equation}
f(t)\propto\frac{\partial_t\Gamma(t)}{(\partial_t\Gamma(t))|_{\Gamma(t)=J}\delta\Gamma(t)},
\end{equation}
which inevitably diverges regardless of the choice of schedules of the transverse field $\Gamma(t)$.

%
%
\subsection{\label{Sec.finite}Finite-size corrections}
In order to deal with the counter-diabatic Hamiltonian at the critical point, we consider finite-size corrections. 
The Holstein-Primakoff transformation up to the third order of the annihilation and the creation operators is given by
\begin{eqnarray}
&S_x=\frac{N}{2}-a^\dag a, \label{Eq.HP1} \\
&S_y=\frac{\sqrt{N}}{2}(a+a^\dag)-\frac{1}{4\sqrt{N}}(a^\dag aa+a^\dag a^\dag a), \label{Eq.HP2} \\
&S_z=\frac{\sqrt{N}}{2i}(a-a^\dag)-\frac{1}{4i\sqrt{N}}(a^\dag aa-a^\dag a^\dag a), \label{Eq.HP3}
\end{eqnarray}
in the disordered phase. 
In the ordered phase, introducing the rotation operator $U^\phi=\exp[-i\phi S_y]$, the spin operators are rotated as
\begin{eqnarray}
&S_x^\phi=U^\phi S_xU^{\phi\dag}=S_x\cos\phi-S_z\sin\phi, \label{Eq.HProtate1} \\
&S_y^\phi=U^\phi S_yU^{\phi\dag}=S_y, \label{Eq.HProtate2} \\
&S_z^\phi=U^\phi S_zU^{\phi\dag}=S_z\cos\phi+S_x\sin\phi. \label{Eq.HProtate3}
\end{eqnarray}
By substituting Eqs.~(\ref{Eq.HP1}), (\ref{Eq.HP2}), and (\ref{Eq.HP3}) for Eq.~(\ref{Eq.BJJ}), performing the harmonic approximation after the normal ordering, and neglecting a constant energy, we obtain
\begin{equation}
\mathcal{H}_0(t)=\frac{J}{2}\left(1-\frac{1}{2N}\right)(a^2+a^{\dag2})+\left[2\Gamma(t)-J\left(1-\frac{1}{N}\right)\right]a^\dag a,
\label{Eq.HamLMGfinitepara}
\end{equation}
in the disordered phase. 
In contrast, in the ordered phase, we perform the same procedure after replacing the spin operators in Eq.~(\ref{Eq.BJJ}) with the rotated spin operators (\ref{Eq.HProtate1}), (\ref{Eq.HProtate2}), and (\ref{Eq.HProtate3}), leading to the Hamiltonian
\begin{eqnarray}
\mathcal{H}_0^{\pm\phi}(t)=&\frac{\Gamma^2(t)}{2J}\left(1-\frac{1}{2N}\right)(a^2+a^{\dag2}) \nonumber \\
&+\left[2J\left(1-\frac{1}{N}\right)-\frac{\Gamma^2(t)}{J}\left(1-\frac{3}{N}\right)\right]a^\dag a \nonumber \\
&\pm\frac{1}{i\sqrt{N}}\frac{\Gamma(t)}{J}\sqrt{J^2-\Gamma^2(t)}(a-a^\dag),
\label{Eq.HamLMGfiniteferro}
\end{eqnarray}
for each well of the potential. 

In order to eliminate the first order terms in the ordered phase, we consider the displaced annihilation and creation operators
\begin{eqnarray}
&a=\tilde{a}+A^\pm, \label{Eq.displace1} \\
&a^\dag=\tilde{a}^\dag-A^\pm, \label{Eq.displace2}
\end{eqnarray}
where $A^\pm$ is a pure imaginary and given by
\begin{equation}
A^\pm=\pm\frac{1}{iN^{1/2}}\frac{\Gamma(t)\sqrt{J^2-\Gamma^2(t)}}{2J^2\left(1-\frac{1}{N}\right)-2\Gamma^2(t)\left(1-\frac{7}{4N}\right)}. 
\end{equation}
Then, the Hamiltonian (\ref{Eq.HamLMGfiniteferro}) is rewritten as
\begin{eqnarray}
\mathcal{H}_0^{\pm\phi}(t)=&\frac{\Gamma^2(t)}{2J}\left(1-\frac{1}{2N}\right)(\tilde{a}^2+\tilde{a}^{\dag2}) \nonumber \\
&+\left[2J\left(1-\frac{1}{N}\right)-\frac{\Gamma^2(t)}{J}\left(1-\frac{3}{N}\right)\right]\tilde{a}^\dag\tilde{a},
\label{Eq.HamLMGfiniteferro2}
\end{eqnarray}
where we again neglected a constant. 

Using the Bogoliubov transformation, we can diagonalize the Hamiltonian (\ref{Eq.HamLMGfinitepara}) and (\ref{Eq.HamLMGfiniteferro2}) in the same way as obtaining Eq.~(\ref{Eq.harmHam}). 
The frequencies are given by
\begin{equation}
\omega(t)=2\left\{\left[\Gamma(t)-\frac{J}{2}\left(1-\frac{1}{N}\right)\right]^2\right.\left.-\left(\frac{J}{2}\right)^2\left(1-\frac{1}{2N}\right)^2\right\}^{1/2},
\end{equation}
in the disordered phase, and given by
\begin{eqnarray}
\omega(t)=&2\left\{\left[J\left(1-\frac{1}{N}\right)-\frac{\Gamma^2(t)}{2J}\left(1-\frac{3}{N}\right)\right]^2\right. \nonumber \\
&\left.-\left(\frac{\Gamma^2(t)}{2J}\right)^2\left(1-\frac{1}{2N}\right)^2\right\}^{1/2},
\end{eqnarray}
for each rotation in the ordered phase. 

For the disordered phase, we can easily go back to the spin operator representation using the relation $b^2-b^{\dag2}=a^2-a^{\dag2}$ and the Holstein-Primakoff transformation up to the second order. 
The counter-diabatic Hamiltonian for the disordered phase is given by Eq.~(\ref{Eq.CDhamLMG}) with
\begin{equation}
f(t)=-\frac{1}{4}\frac{\left[2\Gamma(t)-J\left(1-\frac{1}{N}\right)\right](\partial_t\Gamma(t))}{\left[\Gamma(t)-\frac{J}{2}\left(1-\frac{1}{N}\right)\right]^2-\left(\frac{J}{2}\right)^2\left(1-\frac{1}{2N}\right)^2}. 
\label{Eq.scheDO}
\end{equation}
In contrast, in the ordered phase, we have other terms depending on the directions of rotation due to the displacement (\ref{Eq.displace1}) and (\ref{Eq.displace2}), leading to the relation
\begin{equation}
b^2-b^{\dag2}=\tilde{a}^2-\tilde{a}^{\dag2}=\frac{2i}{N}(S_yS_z+S_zS_y)-\frac{4}{\sqrt{N}}A^\pm S_y.
\end{equation}
However, we sum up both directions of rotation as Eq.~(\ref{Eq.CDhamBS}), and thus the second term proportional to $S_y$ disappears. 
Therefore, the counter-diabatic Hamiltonian for the ordered phase is again given by Eq.~(\ref{Eq.CDhamLMG}) with 
\begin{equation}
f(t)=\frac{\left[\Gamma(t)\left(1-\frac{1}{N}\right)\left(1-\frac{3}{N}\right)+\frac{5\Gamma^3(t)}{2J^2}\left(\frac{1}{N}-\frac{7}{4N^2}\right)\right](\partial_t\Gamma(t))}{\left[J\left(1-\frac{1}{N}\right)-\frac{\Gamma^2(t)}{2J}\left(1-\frac{3}{N}\right)\right]^2-\left(\frac{\Gamma^2(t)}{2J}\right)^2\left(1-\frac{1}{2N}\right)^2}.
\label{Eq.scheO}
\end{equation}
Note that we can easily confirm that the schedule of the counter-diabatic driving does not diverge and is continuous with appropriate choices of schedules of the transverse field when $N<\infty$. 
Of course, both schedules of the counter-diabatic driving (\ref{Eq.scheDO}) and (\ref{Eq.scheO}) converge to Eqs.~(\ref{Eq.scheDO1}) and (\ref{Eq.scheO1}) in the thermodynamic limit $N\to\infty$, respectively. 

%
%
\section{\label{Sec.cat}Cat-state generation}
In this section, we demonstrate generation of the cat state via shortcuts to adiabaticity. 
In order to confirm usefulness of our method, we compare dynamics 
\begin{equation}
i\partial_t|\Psi(t)\rangle=\mathcal{H}(t)|\Psi(t)\rangle,
\end{equation}
with and without the counter-diabatic driving. 
Here, we consider a polynomial schedule
\begin{eqnarray}
\Gamma(t)&=&J[48(t/t_f)^5-120(t/t_f)^4+100(t/t_f)^3-30(t/t_f)^2+2] \label{Eq.scheGamma1} \\
&=&J[48s^5-120s^4+100s^3-30s^2+2], \label{Eq.scheGamma2}
\end{eqnarray}
where $s=t/t_f$. 
The coefficients of the schedule are determined from the boundary conditions $\Gamma(0)=2J$, $\Gamma(t_f/2)=J$, and $\Gamma(t_f)=0$ and $\partial_t\Gamma(t)|_{t=0,t_f/2,t_f}=0$. 
This setup leads to the continuous counter-diabatic driving which vanishes at the initial and the final time. 
Throughout this paper, we put $J=1$. 

\begin{figure}
\centering
\includegraphics[width=10cm]{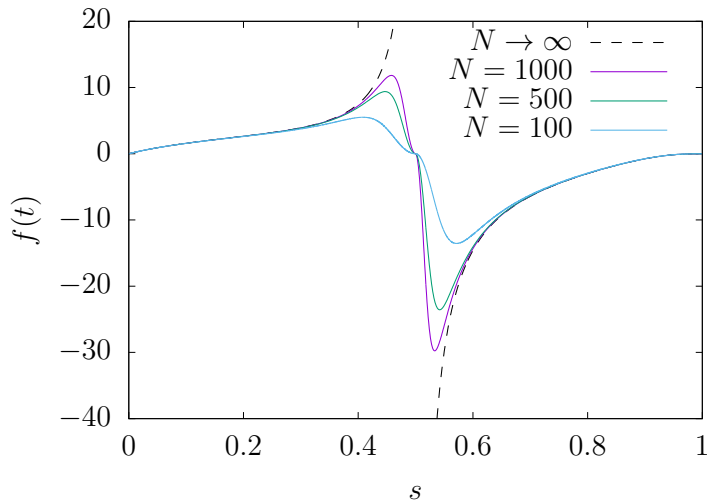}
\caption{\label{Fig.CDfields}Schedules of the counter-diabatic driving for $N=100$, $500$, $1000$, and $\infty$. Here, $t_f=1$. }
\end{figure}

The schedules of the counter-diabatic driving for $N=100$, $500$, $1000$, and the thermodynamic limit $N\to\infty$ with the transverse field (\ref{Eq.scheGamma1}) are plotted in Fig.~\ref{Fig.CDfields}. 
As mentioned in the previous works~\cite{PhysRevE.87.062117,PhysRevLett.114.177206} and above, the counter-diabatic driving diverges in the thermodynamic limit $N\to\infty$ regardless of schedules. 
In contrast, the schedules of the counter-diabatic driving (\ref{Eq.scheDO}) and (\ref{Eq.scheO}) for finite $N$ are continuous and become zero at the critical point. 

%
%
\subsection{Fidelity to the cat state}
\begin{figure}
\centering
\includegraphics[width=10cm]{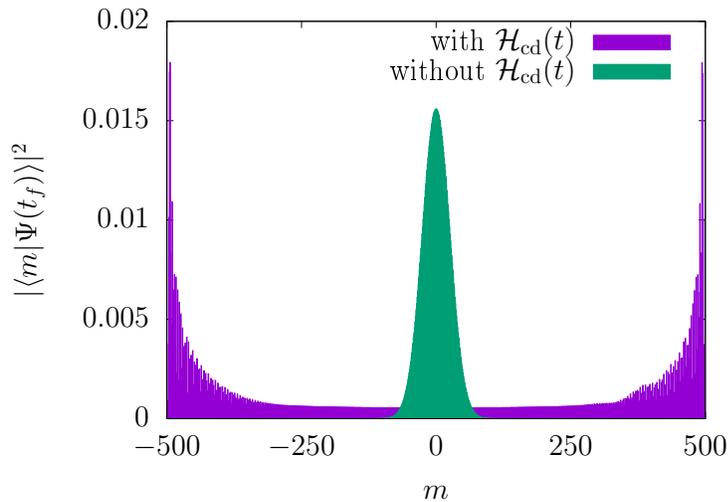}
\caption{\label{Fig.dist}Distributions of the populations at the final time $t_f$ for $N=1000$ and $t_f=1$. }
\end{figure}

We confirm that our method can improve fidelity to the cat state. 
For this purpose, we compare the distributions of the eigenstate populations at the final time, i.e., the distributions of $|\langle m|\Psi(t_f)\rangle|^2$, where $|m\rangle$ is the eigenstate of $S_z$, with and without the counter-diabatic Hamiltonian. 
From the Hamiltonian (\ref{Eq.BJJ}), it is obvious that the low-energy states are given by large $|m|$ and the high-energy states have small $|m|$. 
If the perfect adiabatic dynamics is attained, the distribution consists of two peaks at $m=\pm N/2$ with the populations $0.5$, respectively. 
We plot the distributions for $N=1000$ in Fig.~\ref{Fig.dist}. 
With the counter-diabatic Hamiltonian, the state is distributed over low-energy states and forms a cat state. 
In contrast, without the counter-diabatic Hamiltonian, the state is distributed over high-energy states. 
Note that even with the counter-diabatic Hamiltonian, there is a small amount of populations in unfavorable high-energy states. 
This unfavorable excitations are due to the approximation and fast operation. 
Slower operation can suppress this unfavorable excitations but deviations due to the approximation are accumulated. 
Therefore, we have to optimize operation time. 

\begin{figure}
\centering
\includegraphics[width=10cm]{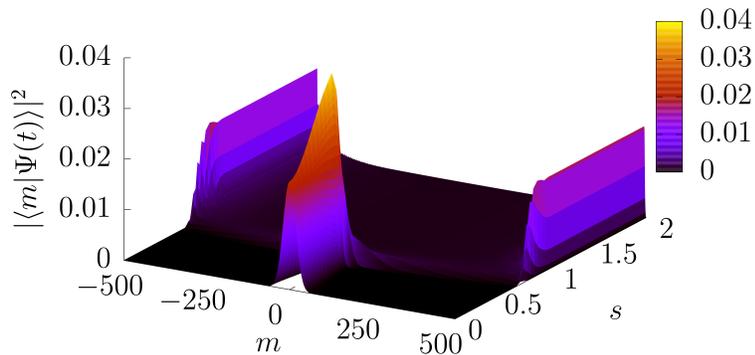}
\caption{\label{Fig.long-dist}Time-evolution of the distribution of the eigenstate populations for $N=1000$ and $t_f=1$. }
\end{figure}

As mentioned in Sec.~\ref{Sec.Intro}, automatically freezing dynamics after creation of cat states is one of the advantages of adiabatic generation. 
We demonstrate generation of the cat state within time $s=1$ and dynamics with the final Hamiltonian $\mathcal{H}(t_f)$ up to time $s=2$, and plot in Fig.~\ref{Fig.long-dist}. 
It is obvious that the cat state is maintained after the creation process and thus we do not have to care about timing unless decoherence becomes problematic. 

\begin{figure}
\centering
\includegraphics[width=10cm]{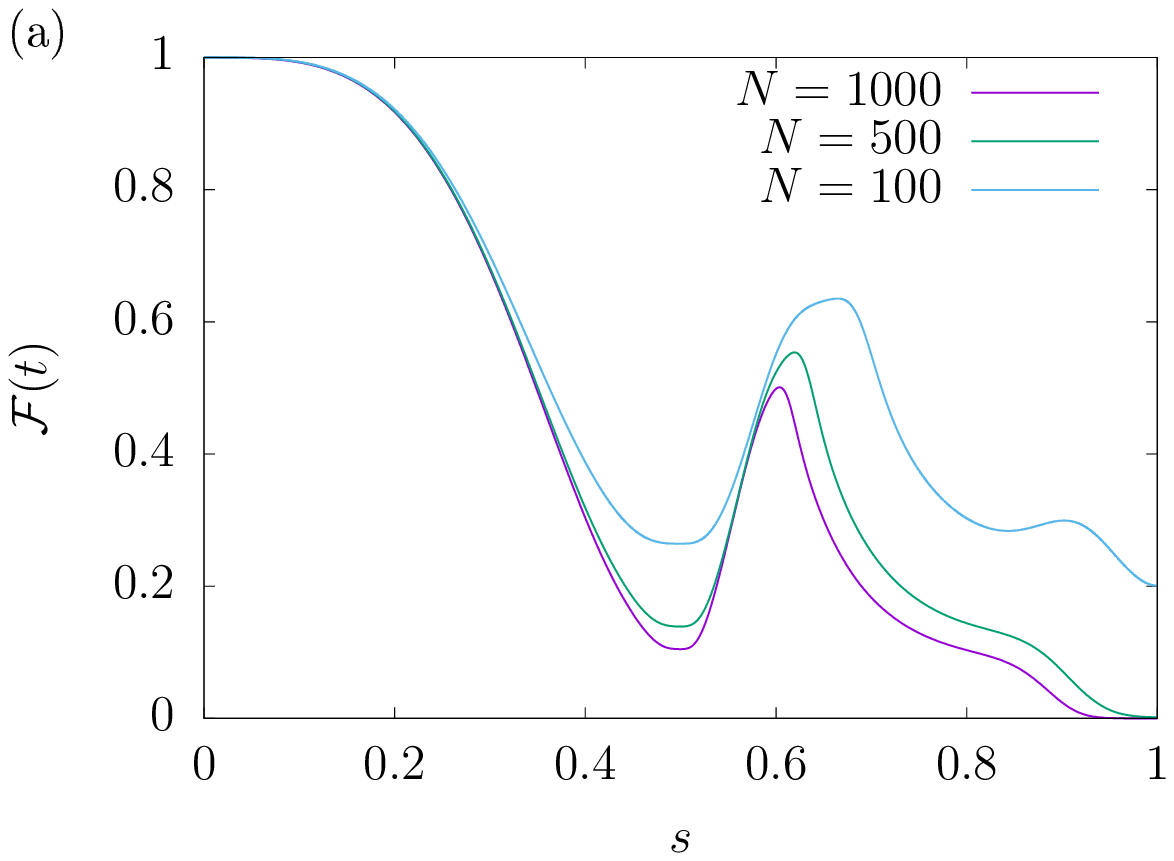}
\includegraphics[width=10cm]{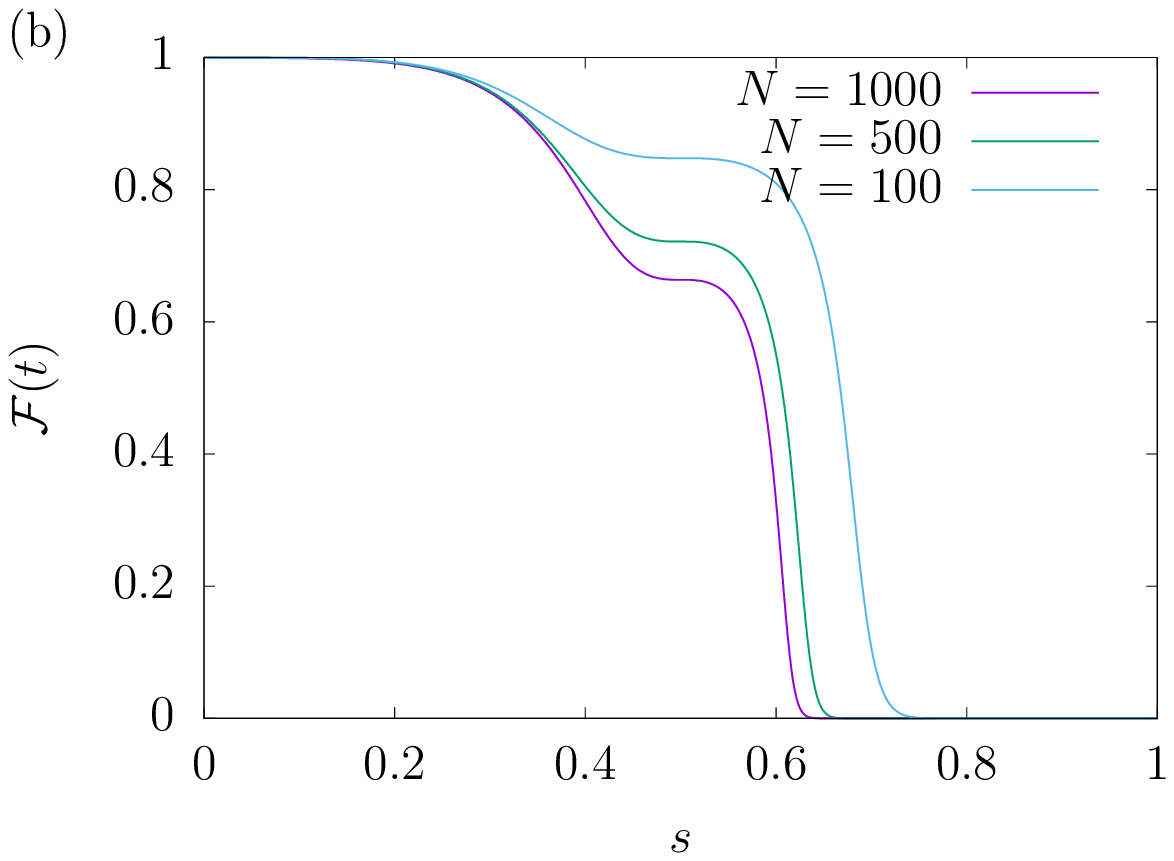}
\caption{\label{Fig.fid}Fidelity to the ground-state subspace for $N=100$, $500$, and $1000$ (a) with and (b) without the counter-diabatic Hamiltonian. Here, $t_f=1$. }
\end{figure}

Now, we consider the fidelity to the ground state subspace $\mathcal{F}(t)=|\langle\Psi_\mathrm{GS}(t)|\Psi(t)\rangle|^2$, where $|\Psi_\mathrm{GS}(t)\rangle$ is the ground state for the disordered phase, $\Gamma(t)>J$, and is the sum of the ground state and the first excited state for the ordered phase, $\Gamma(t)<J$. 
As discussed in Ref.~\cite{doi:10.7566/JPSJ.86.094002}, this criterion of adiabaticity is rather severe especially for large systems $N\to\infty$. 
This is because not only deviations are enhanced but also the gap closes in the limit $N\to\infty$. 
We plot the fidelity $\mathcal{F}(t)$ of both cases, with and without the counter-diabatic Hamiltonian, in Fig.~\ref{Fig.fid}. 
Although it is the severe criterion, the fidelity to the ground state subspace remains finite for $N=100$. 

It should be noted that the counter-diabatic Hamiltonian results in bad fidelities to the ground state subspace around the critical point (compare Fig.~\ref{Fig.fid} (a) and (b) around $s=1/2$), which was also observed in the previous work (see Supplemental Material of Ref.~\cite{PhysRevLett.114.177206}). 
This should be attributable to the breakdown of the harmonic approximation as discussed in Sec.~\ref{Sec.BJJ}. 
One might expect that we can improve adiabaticity by turning off the counter-diabatic driving up to the critical point and turning on it at there. 
However, at last, results are almost the same. 

%
%
\subsection{Residual energy and incomplete magnetization}
It is also of interest how the counter-diabatic driving can accelerate adiabatic dynamics. 
For this purpose, we calculate the residual energy and the incomplete magnetization. 

The residual energy is defined as the deviation of the energy from the ground state energy at the final time, $t=t_f$,
\begin{equation}
E_\mathrm{res}=E(t_f)-E_\mathrm{gs}, 
\end{equation}
where $E(t_f)$ is the energy given by the dynamical state, $E(t_f)=\langle\Psi(t_f)|\mathcal{H}(t_f)|\Psi(t_f)\rangle$, and $E_\mathrm{gs}$ is the ground state energy, $E_\mathrm{gs}=-JN/2$. 
The incomplete magnetization is given by the deviation of the order parameter, which is given by
\begin{equation}
m(t_f)=\sqrt{\frac{1}{S^2}\langle\Psi(t_f)|S^{z2}|\Psi(t_f)\rangle},
\end{equation}
from that of the ground state 
\begin{equation}
m_\mathrm{inc}=m_\mathrm{gs}-m(t_f),
\end{equation}
where $m_\mathrm{gs}$ is the order parameter of the ground state given by $m_\mathrm{gs}=1$. 

\begin{figure}
\centering
\includegraphics[width=10cm]{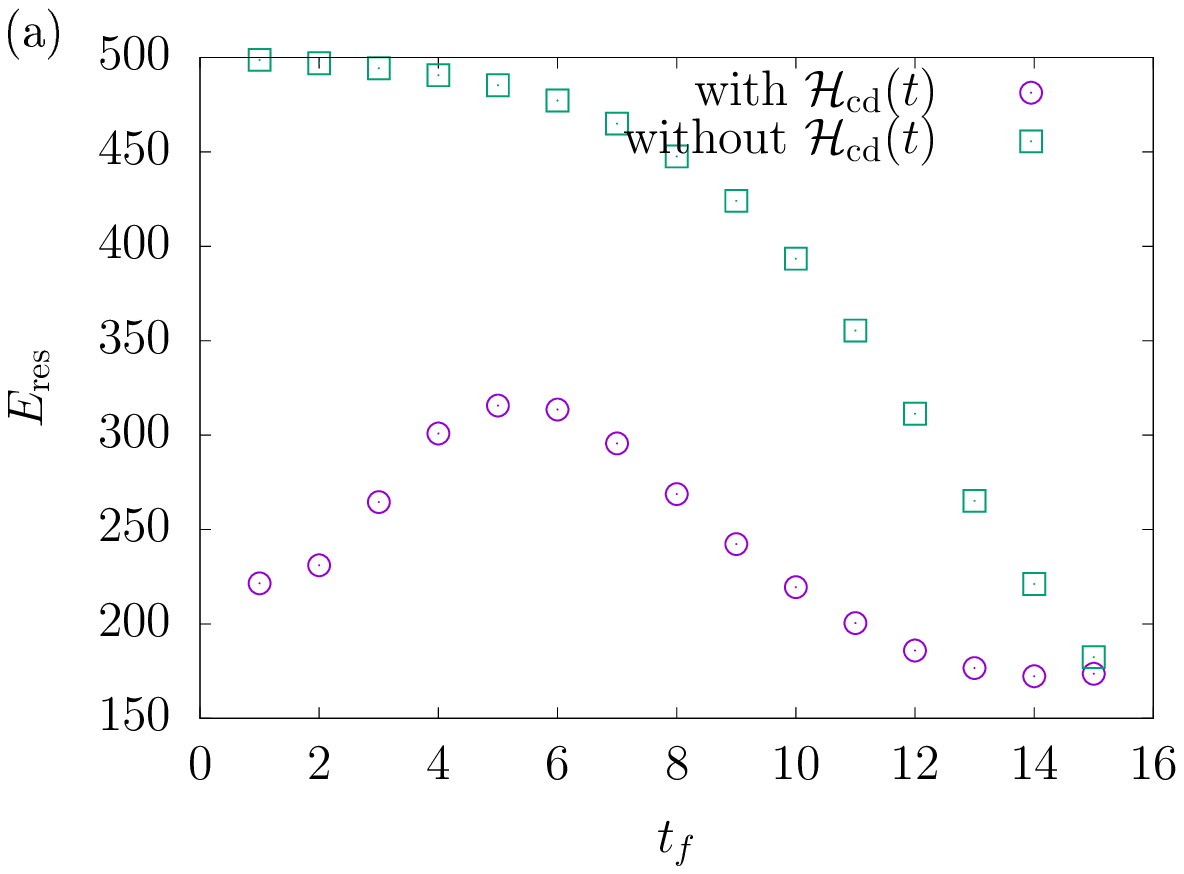}
\includegraphics[width=10cm]{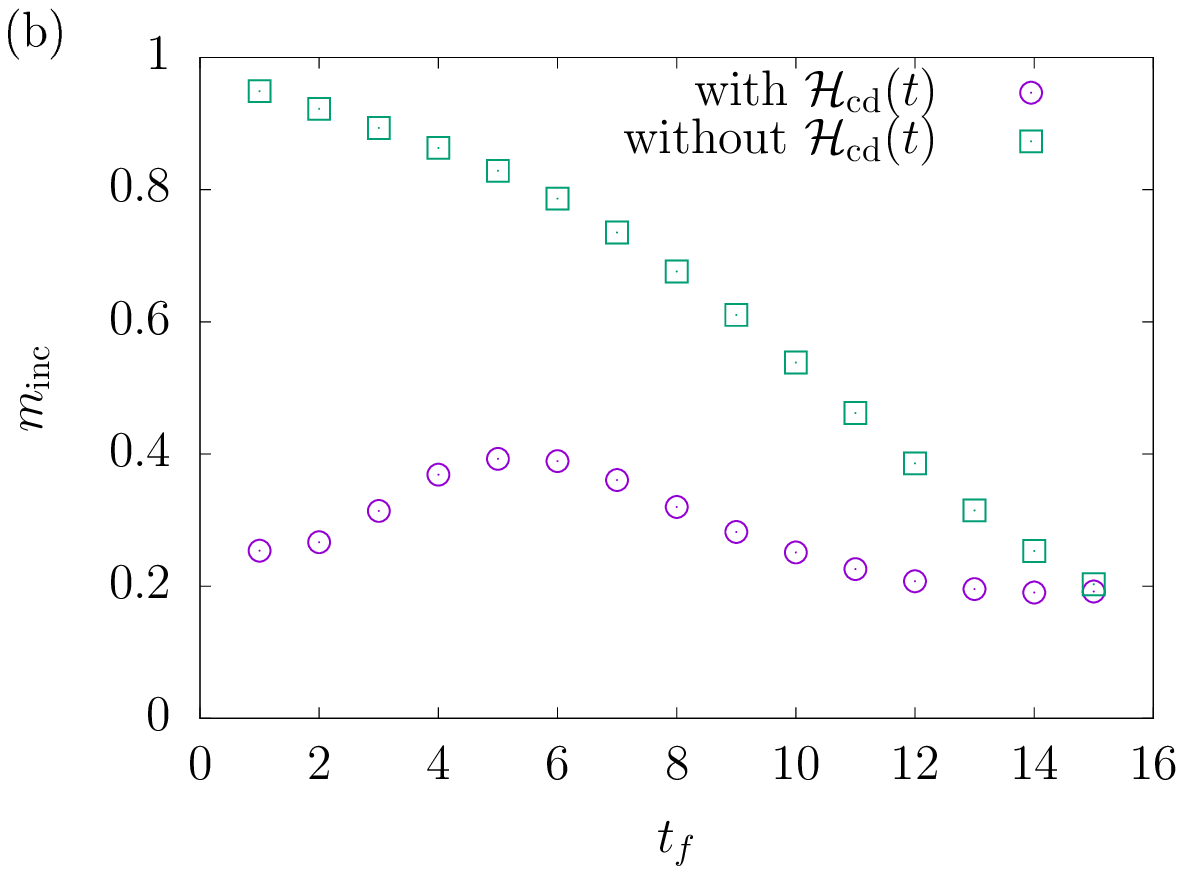}
\caption{\label{Fig.residual}(a) Residual energy and (b) incomplete magnetization for $N=1000$ with and without the counter-diabatic Hamiltonian. }
\end{figure}

We compare the residual energy and the incomplete magnetization in the cases with and without the counter-diabatic Hamiltonian plotted in Fig.~\ref{Fig.residual}. 
For fast operations, where bare adiabatic tracking results in failure, we can suppress diabatic transitions using the counter-diabatic driving. 
Roughly speaking, the counter-diabatic driving accelerates dynamics by ten times or more compared with adiabatic tracking. 
Note that this approximated counter-diabatic driving leads to rather bad results when bare adiabatic tracking gives enough adiabaticity. 
However, this does not lower worth of our results because acceleration of adiabatic dynamics is not necessary if the dynamics is already adiabatic. 

%
%
\subsection{Quantum Fisher information}
Now, we estimate the quantum Fisher information $F_Q$ of the state driven by the counter-diabatic Hamiltonian. 
If the quantum Fisher information satisfies $F_Q>N$, then we can conclude that the state is entangled and has phase sensitivity beyond the standard quantum limit, i.e., potentially useful applying to quantum metrology. 
The upper bound of the quantum Fisher information is given by $F_Q=N^2$, and then phase sensitivity reaches the Heisenberg limit. 
Note that, for examples, the quantum Fisher information of the GHZ state and the NOON state is given by $F_Q=N^2$, which reaches the Heisenberg limit, and that of the maximally spin-squeezed Dicke state is given by $F_Q=N^2/2+N$, which is beyond the standard quantum limit but does not reach the Heisenberg limit. 

\begin{figure}
\centering
\includegraphics[width=10cm]{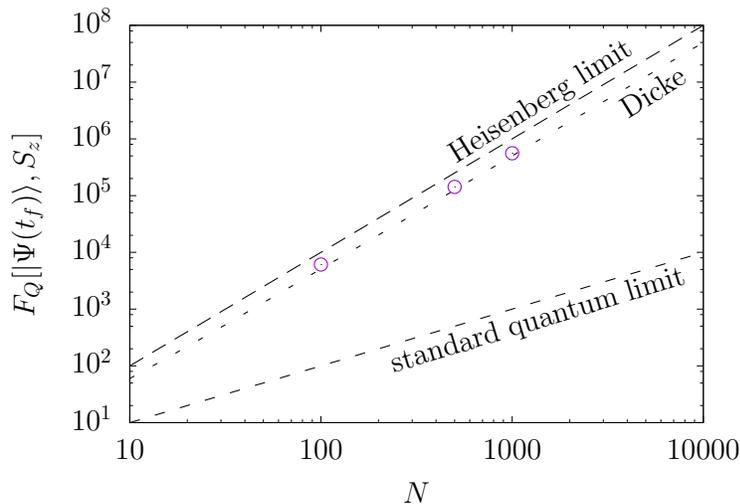}
\caption{\label{Fig.fisher}Quantum Fisher information of the cat states generated by the counter-diabatic driving for $N=100$, $500$, and $1000$. Here, $t_f=1$. }
\end{figure}

In our scheme, the state is the pure state and the relative Hermitian operator is $S_z$, and thus the appropriate quantum Fisher information is given by
\begin{equation}
F_Q[|\Psi(t_f)\rangle,S_z]=4(\langle\Psi(t_f)|S_z^2|\Psi(t_f)\rangle-\langle\Psi(t_f)|S_z|\Psi(t_f)\rangle^2).
\end{equation}
We plot the quantum Fisher information for $N=100$, $500$, and $1000$ with $t_f=1$ in Fig.~\ref{Fig.fisher}. 
The results show that the quantum Fisher information is far beyond the standard quantum limit, slightly above that of the maximally spin-squeezed Dicke state, and of course below the Heisenberg limit. 
It is evident that the generated cat states are truly macroscopic entangled states. 
It is also remarkable that the quantum Fisher information of the generated cat states is larger than that of the maximally spin-squeezed Dicke state, which is the ground state of the anti-ferromagnetic Lipkin-Meshkov-Glick model, and thus our scheme of quantum speedup is significantly useful. 

%
%
\section{\label{Sec.conc}Conclusion remarks}
We proposed the quantum speedup method to create the cat state in bosonic Josephson junctions using the counter-diabatic driving. 
By taking finite-size corrections into account, the problem of divergence in counter-diabatic driving was avoided. 
The cat state was successfully produced via shortcuts. 
In order to confirm usefulness of our method, we calculated the distribution of the eigenstate populations, the fidelity to the ground state subspace, the residual energy, the incomplete magnetization, and the quantum Fisher information. 
The results of the residual energy and the incomplete magnetization suggest that our scheme mimics adiabatic dynamics faster than bare adiabatic tracking by ten times or more. 
As to the quantum Fisher information, that of the cat state generated by our scheme exceeded that of the maximally spin squeezed Dicke state, which strongly supports usefulness of our method. 

Our scheme should be feasible in experiments using a Bose-Einstein condensate in a toroidal trap, where the scheme implementing interactions in the forms of the Hamiltonian (\ref{Eq.BJJ}) and of the counter-diabatic Hamiltonian (\ref{Eq.CDhamLMG}) has been theoretically proposed~\cite{PhysRevA.91.053612}, although it has not been observed in experiments yet. 
It would be more implementable if we can simplify the form of the Hamiltonian by adequately designing shortcuts~\cite{PhysRevE.87.062117,PhysRevA.95.062319}. 

There are some possible candidates which further improve our method. 
One of the methods is to control the parameters according to the schedules satisfying the fixed-point condition, where improvement of fidelity to the ground state was observed~\cite{PhysRevE.87.062117}. 
Combining with the optimization approach and considering higher order terms of counter-diabatic driving are, of course, able to improve adiabaticity~\cite{PhysRevLett.114.177206}. 
As to optimization, our schedule of the counter-diabatic driving should be guidance to find optimized schedules. 

Finally, we make reference to other systems, where our method might be feasible. 
One of the candidates is circuit QED systems, where the scheme to realize the Lipkin-Meshkov-Glick model was recently proposed~\cite{0295-5075-90-5-54001}. 
Single-molecular magnets are also candidates for our method, where counter-diabatic terms can be controlled by pressures~\cite{ADFM:ADFM200500244}. 

%
%
\ack
The author is grateful to Professor Seiji Miyashita and Dr. Takashi Mori for useful comments on the manuscript. 
This work is supported by the Ministry of Education, Culture, Sports, Science and Technology (MEXT) of Japan through the Elements Strategy Initiative Center for Magnetic Materials. 
The author is supported by the Japan Society for the Promotion of Science (JSPS) through the Program for Leading Graduate Schools: Material Education program for the future leaders in Research, Industry, and Technology (MERIT) of the University of Tokyo. 

%
%
\section*{References}
\bibliographystyle{iopart-num}
\bibliography{BECcat_NJPbib}

\end{document}